\documentclass[fleqn,twoside]{article}
\usepackage{gc}
\usepackage{amssymb}

\heads{Vladimir V. Kassandrov}
      {A non-stationary generalization of the Kerr congruence}

\newcommand{\be}{\begin{equation}\label}
\newcommand{\ee}{\end{equation}}
\newcommand{\prt}{\partial}
\newcommand{\p}{\prime}
\newcommand{\bib}{\bibitem}
\newcommand{\nabl}{{\bf \nabla}}         
\newcommand{\wh}{\widehat}

\begin{document}
\twocolumn[

\Title{A NON-STATIONARY GENERALIZATION        \yy
       OF THE KERR CONGRUENCE}


   \Author{Vladimir V. Kassandrov\foom 1            
              }   
          {Institute of Gravitation and Cosmology, Peoples' Friendship University of Russia, \\
Miklukho-Maklaya St. 6, Moscow, 117198 Russia}              


\Abstract
    {Making use of the Kerr theorem for shear-free null congruences and of Newman's   
representation for a virtual charge ``moving'' in complex space-time, we obtain 
an axisymmetric time-dependent generalization of the Kerr congruence, with a 
singular ring uniformly contracting to a point and expanding then to infinity. 
Electromagnetic and complex eikonal field distributions are naturally 
associated with the obtained congruence, with electric charge being necesssarily  
unit (``elementary''). We conjecture that the corresponding solution to the 
Einstein-Maxwell equations could describe the process of continious transition 
of the naked ringlike singularitiy into a rotating black hole and vice versa, 
under a particular current radius of the singular ring.} 

]  
\email 1 {vkassan@sci.pfu.edu.ru}

\section{Shear-free null congruences and associated symmetries and fields}

A broad class of physically relevant solutions to the Einstein or 
Einstein-Maxwell equations is generated by a shear-free congruence of 
null rays. In particular, the Schwarzchild (Reissner-Nordstr\"om) and Kerr 
(Kerr-Newman) solutions belong to this class. Since the times of 
the paper by Debney, Kerr and Schild~\cite{Debney},  
it is known that the Einstein equations are partly satisfied after  
substitution of the Riemannian metric 
\be{schild}
g_{\mu\nu} = \eta_{\mu\nu} + h k_\mu k_\nu,
\ee
with a null 4-vector field $k:~k_\mu k^\mu =0$ obeying the defining equations 
of {\it shear-free null congruences} (SFNC) in Minkowski background space with 
the metric $\eta_{\mu\nu}$ (as a consequence, these are automatically geodetic, 
composed of rectilinear rays). In the above-mentioned cases it is, moreover, 
possible to choose the scalar ``gravitational potential'' $h$ in such a way 
that the rest of equations are fulfilled as well. Note that properties of 
a congruence to be null, geodetic and shear-free are preserved 
by themselves under the {\it Kerr-Schild\/} deformation (\ref{schild}) of 
the Minkowski metric. 

On the other hand, the SFNC equations may be represented in a spinorial  
form~\cite{Penrose}
\be{sfc}
\xi^A \xi^B \nabl_{B^\p B} \xi_A = 0 
\ee 
for a principal 2-spinor field of the congruence $\xi=\{\xi_A\},~A=1,2$ 
up to a phase factor defined by the null 4-vector field $k_\mu = \xi_A\xi_{A^\p}$.         
It has been in fact shown in the very work~\cite{Debney} that, in the Minkowski 
background, SFNCs carry a {\it twistorial structure\/} corresponding to the 
principal spinor $\xi$ and, as a result, they may be completely described in a 
purely algebraic way. Such a remarkable statement is known as the 
{\it Kerr theorem\/}~\cite{Penrose} and, for our further purposes, 
can be formulated in an invariant form~\cite{Joseph,Rev}.

Specifically, consider an arbitrary and (almost everywhere) analytical 2-surface in 
the 4-dimensional complex (twistor) space ${\bf W}=\{\xi,\tau\}$. The surface 
is represented implicitly through a set of two algebraic constraints
\be{constr}
\Pi^C(\xi,\tau) = 0,~~~C=1,2,
\ee
so that only two of the four twistor components are functionally independent. 

Let us now make use of the twistor-defining incidence relation~\cite{Penrose} 
\be{inc}
\tau = X \xi, ~~~(\tau^{A^\p} = X^{A^\p A}\xi_A),
\ee
linking the two spinor constituents $\{\xi,\tau\}$ with the points $X$ of 
the Minkowski space $\bf M$ represented by a $2\times 2$ {\it Hermitean} matrix $X=X^+$.  
After substitution of (\ref{inc}) into (\ref{constr}) the latter turns to be  
a pair of algebraic equations 
\be{kerr}
\Pi^C(\xi,X\xi) = 0,~~~C=1,2,
\ee
for two unknowns $\xi$, with the coordinates $\{X^{A^\p A}\}$ playing the 
role of parameters. Successively resolving (\ref{kerr}) at different 
points $X\in \bf M$, one obtains a {\it multivalued} solution $\xi(X)$ 
whose every branch defines, in fact, a SFNC (exactly, a {\it sub-congruence} 
of this type). 

Moreover, the spinor found obeys an overdetermined  set of four equations
\be{sfcN}
\xi^B \nabl_{B^\p B}\xi_A = 0, 
\ee 
from which the standard form of the SFNC equations (\ref{constr}) results 
through contraction with $\xi^A$. 

To prove this, one differentiates (\ref{kerr}) in each of the coordinates 
$X^{B^\p B}$ to get
\be{diff}
Q^{CA} \nabl_{B^\p B} \xi_A = -\frac{\prt \Pi^C}{\prt \tau^{B^\p}} \xi_B,
\ee
with the $2\times 2$ matrix $Q^{CA}$ of the following form: 
\be{qmat}
Q^{CA} = \frac{d\Pi^C}{d\xi_A} = \frac{\prt \Pi^C}{\prt \xi_A} +
\frac{\prt \Pi^C}{\prt \tau^{A^\p}} X^{A^\p A}.
\ee
At the points defined by the condition 
\be{caust}
\det \Vert Q^{CA} \Vert = 0,
\ee
the derivatives of the principal spinor are indefinite, so that (\ref{caust})  
turns out to be the condition of a {\it caustic locus} for rays of the SFNC. 
On the other hand, at all regular points contracting (\ref{diff}) with $\xi_B$
one gets the afore-presented set of equations (\ref{sfcN}).

One can furthermore notice that, with respect to the {\it ratio} of components 
of the principal spinor $\xi$, Eqs. (\ref{sfc}) and (\ref{sfcN}) are 
entirely equivalent, the latter only specifying, in addition, {\it each\/} 
of the two spinor components. This is in correspondence with the symmetries of 
these equations. 

Indeed, the standard SFNC equations (\ref{sfc}) as well as the incidence 
relation (\ref{inc}) preserve their form under arbitrary rescaling 
$\xi \mapsto \alpha(X) \xi$, and one can easily show that the gauge parameter 
$\alpha$ may be chosen in such a way that the transformed spinor satisfies 
the stronger equations (\ref{sfcN}). The latter possess a restricted 
reparametrization invariance~\cite{Joseph,Rev}, with the parameter $\alpha$ 
depending on the coordinates only {\it implicitly}, via the components of the 
transforming spinor or its twistor counterparts, $\alpha=\alpha(\xi,\tau)$. 
Evidently, this restricted symmetry is 
also a symmetry of the algebraic constraint (\ref{kerr}) and, geometrically, 
it has the meaning of an arbitrary {\it diffeomorphism in twistor space\/} 
$\{\xi,\tau\}$. For the associated electromagnetic field (see below) it 
manifests itself as a ``weak'' version of the familiar {\it gauge symmetry}.

Any SFNC allows for definition of a (complexified) electromagnetic-like field
~\cite{AD,GR95}. Specifically, making use of spinor 
algebra, one can write down the SFNC equations (\ref{sfc}) in the   
equivalent form~\cite{Torres,Trish} 
\be{pot}
\nabl_{B^\p (B} \xi_{A)} = \Phi_{B^\p (B}\xi_{A)}
\ee
where some complex 4-vector $\Phi_{B^\p B}$ comes into play, and the parentheses 
denote symmetrization in spinor indices. Analogously, Eqs. (\ref{sfcN}) 
read
\be{potN} 
\nabl_{B^\p B} \xi_A = \Phi_{B^\p A}\xi_B 
\ee
(with the skew-symmetric part being defined in addition to (\ref{pot})). 
Under rescalings of 
the principal spinor, the 4-vector $\Phi_{B^\p B}$ transforms gradient-wise and 
can thus be interpreted as the potential of an Abelian complex gauge field. 
As a consequence of the integrability conditions for (\ref{pot}), the field 
strengths 
\be{strength}
\varphi_{(AB)}=\nabl_{B^\p (B} \Phi^{B^\p}_{A)} 
\ee
are anti-self-dual~\cite{GR95,Asya} 
\be{selfdual}
\nabl_{(A^\p B} \Phi_{B^\p)}^B = 0
\ee
and thus satisfy the homogeneous Maxwell equations
\be{maxwell}
\nabla_{B^\p}^A ~\varphi_{(AB)} = 0.
\ee 
 
The field strengths are rescaling (gauge-) invariant 
and can be expressed 
through second-order derivatives of the only complex function $g(X)$,  
the ratio of two complex components, say, $g=\xi_{2^\p}/\xi_{1^\p}$~\cite{Asya}.   
With respect to the {\it projective spinor} component $g(X)$, the generating  
constraints (\ref{kerr}) reduce to the only one 
\be{reduce}       
\Pi(g,\tau^1,\tau^2)=0,
\ee
containing three {\it projective\/} twistor components. Introducing a canonical 
representation  of coordinates on $\bf M$, 
\be{repr}
X=X^+ =\left(
\begin{array}{cc}
u & w \\
p & v
\end{array}\right)
=\left(
\begin{array}{cc}
ct+z & x-\imath y \\
x+\imath y & ct-z
\end{array}\right),
\ee
with $u,v$ being real and $p,w$ complex-conjugated, the incidence relation 
(\ref{inc}) takes the form of two equations:
\be{incg}
\tau^1 = u+wg,~~\tau^2 = p+vg,
\ee
so that the constraint (\ref{reduce}) becomes an equation for the sole 
unknown $g$,
\be{reduce2}
\Pi(g, u + wg, p + vg)=0,
\ee
the condition (\ref{caust}) for SFNC singularities (caustics) simplifies to 
\be{sing}
P:=\frac{d\Pi}{dg}\equiv \frac{\prt \Pi}{\prt g} + w \Pi_1 + v\Pi_2 = 0,   
\ee
and the field strengths of the associated Maxwell field (\ref{strength}) 
can be expressed explicitly through the (first and second order) derivatives 
$\Pi_A, \Pi_{AB}$ of generating function $\Pi$ with respect to its twistor 
arguments $\tau^A$~\cite{Joseph,Rev}:
\be{strtwist}
\varphi_{(AB)}=\frac{1}{P}\left(\Pi_{AB} - \frac{d}{dg}\{\frac{\Pi_A \Pi_B}
{P}\}\right). 
\ee
Note that, in the gauge used, from the sole constraint (\ref{reduce2}) 
one obtains a projective spinor component $g(X)$ that defines a SFNC. 
It is, in fact, the content of the {\it Kerr theorem\/} that {\it any} 
SFNC on $\bf M$ may be obtained in such an algebraic way. In a gauge-invariant 
form, the same is true if one starts from the pair of algebraic 
constraints (\ref{constr}). 

Remarkably, the ratio $g(X)$ of spinor components for any SFNC (which may be 
obtained from (\ref{reduce2}) with some generating function $\Pi$) identically 
satisfies the two fundamental Lorentz-invariant equations, both together, namely, 
the linear wave equation $\nabl^2 g =0$ and the nonlinear eikonal 
equation $(\nabla g)^2 = 0$. In a spinorial gauge-invariant form, the 
corresponding relations have been written down in~\cite{Trish}.

As to the introduced Maxwell field, it is exceptional in a number of properties. 
Contrary to some other types of Maxwell fields related to SFNC and introduced some 
time ago, say, in~\cite{Robin,Penrose,Torres,Lind}, the above introduced 
field is gauge-invariant, generally non-null and does not require explicit 
integration of Maxwell's equations themselves. Though this field 
{\it almost everywhere} obeys the {\it homogeneous\/} (``free'') 
Maxwell equations, it nonetheless defines some effective singular ``sources'' 
located at the caustic points (\ref{sing}) of the congruence. Remarkably, all 
these solutions to Maxwell equations are generated by {\it arbitrary} 
twistor functions $\Pi({\bf W})$ and can be obtained in a purely algebraic way, 
via resolving (\ref{reduce2}) and successive calculation of the strengths 
({\ref{strtwist}).

On the other hand, only a distinguished subclass of solutions to the 
Maxwell's equations may be obtained in such a way, and for this reason a sort of 
``selection rules'' naturally arise. In particular, for any SFNC any 
(bound in 3-space) singularity of the electric field (\ref{strtwist}),    
defined by the condition (\ref{sing}), is either neutral or {\it carries an 
electric charge multiple to a minimal ``elementary'' charge} possible 
in the model presented. The ``elementary'' charge  arises as that of a 
ringlike singularity of the fundamental Kerr-like solution considered below. 
The general theorem of charge quantization 
proved in~\cite{sing04} makes use of considerations resembling those in 
Dirac's theory of magnetic monopoles.

It is noteworthy that, in addition to the eikonal, Maxwell and metric (curvature)  
fields, it is possible to associate with any SFNC (i.e., with any generating 
twistor function) a pair of 2-spinors, each satisfying the massless Weyl 
equations~\cite{sing04} and the (``restricted'' $SL(2,\mathbb{C})$) Yang-Mills type 
gauge field equations~\cite{GR95,Asya} (see a discussion in ~\cite{Joseph}).   
We thus see that a shear-free null congruence (exactly, a closed system of 
sub-congruences generating by a unique complex structure) in a Minkowski or  
Kerr-Schild background is really a fundamental geometrophysical object giving 
itself rise to a whole system of massless relativistic fields and singular 
``particlelike'' formations, the latter revealing some properties of real 
quantum particles. In the {\it algebrodynamical} paradigm, as it has been 
demonstrated in~\cite{AD,GR95,Rev}, the naturally formulating conditions of 
{\it biquaternionic analyticity} (``generalized Cauchy-Riemann equations'') 
are in fact equivalent to the defining equations of SFNC.

\section{Newman's ``virtual charge'' and the complex null cone representation}

Shear-free congruences of null rays are especially important because it is 
these congruences that are generated by an arbitrary moving charge through its 
local light cone. 
Making use of such a congruence, for simplest cases one can integrate the Einstein or 
Einstein-Maxwell set of equations and thus associate the electromagnetic and 
metric (curvature) field with a moving pointlike charge/mass~\cite{Kinnersley,Gursey}. 
Newman et al.~\cite{Lind,Newman} generalized this procedure by consideration of 
the complex extension ${\mathbb C}{\bf M}$ of the Minkowski space $\bf M$. 
A ``virtual'' pointlike singularity ``moving'' therein along a ``trajectory'' 
\be{traj}
z^0_\mu (\sigma), ~~\sigma \in {\mathbb C}, ~~~\mu=0,1,2,3
\ee
generates the complexified Lienard-Wiechert electomagnetic and gravitational 
fields obtained by integration of the correspondent field equations and 
subsequent restriction of the solutions onto the real space-time ``cut'' $\bf M$.  
	
On the other hand, one can suggest an alternative, purely algebraic and direct 
method of generating particular solutions for, say, Maxwell and 
eikonal fields described in the previous section: these  correspond 
to  Newman's ``virtual'' pointlike singularity in complex space. 
Specifically, consider the case of the generating function (\ref{reduce}) such 
that it is identically nullified by the 1-parametric substitution 
\be{subs}
z_\mu = z^0_\mu (\sigma)
\ee
(recall that we are now in the $\mathbb{C}\bf M$-space with 
{\it complex} coordinates $z_\mu$). If this is true, then, on  
the complex ``trajectory'' (\ref{traj}), the twistor field is {\it indefinite}, 
and the world line (\ref{subs}) is in fact a {\it focal curve} of the 
congruence. All such congruences can be generated from the 
corresponding relations for twistor components
\be{rel}
\begin{array}{cc}
\tau^1 = u + wg & =u^0(\sigma) + w^0(\sigma) g, \\
\tau^2 = p + vg & = p^0(\sigma) + v^0(\sigma) g,
\end{array}
\ee
where a representation of coordinates similar to (\ref{repr}) is used,   
the distinction being only in the current independence of all four 
complex coordinates of virtual charge $u^0,v^0,,p^0,w^0$. 

Eliminating, say, the parameter $\sigma$ from the two 
equations (\ref{rel}),  we return to an algebraic constraint of the form 
(\ref{reduce}) with some generating function $\Pi$ and may thus be sure that 
the arising SFNC will necessarily possess a focal curve, a 
``world line'' of a virtual charge ``moving'' in the complex 
extension $\mathbb{C}\bf M$. Note that for a complex ``trajectory''  
(\ref{traj}) the whole focal curve will, generally, have no  
points of intersection with the real Minkowski cut $\bf M$.  
In fact, on $\bf M$ one can find only the stringlike {\it caustics} 
of the congruence defined by two real (one complex) constraints (\ref{sing}) 
imposed on four Minkowski coordinates.

Equivalently, one can eliminate the projective spinor component $g$ from 
the two equations (\ref{rel}) and get, instead of (\ref{reduce2}), the 
following generating equation of the {\it complex null cone}: 
\be{cone}
S(\sigma):=Z^{A^\p A}(\sigma) Z_{A^\p A}(\sigma) = 0,~~ 
\ee
where the {\it relative} coordinates 
\be{relcoord}
Z_{A^\p A}:=x_{A^\p A}-z^0_{A^\p A}(\sigma)
\ee
come into play,  and the points of interest are taken in a real $\bf M$ 
subspace $z_{A^\p A}\mapsto x_{A^\p A}$. For any $X = \{x_{A^\p A}\}$, 
resolving this equation with respect to $\sigma$, one finds a 
number of branches of the field of ``retarded complex time'' $\sigma(X)$,  
any of which possesses the following properties.

	1. As in the real case, the field $\sigma(X)$ satisfies 
the complex eikonal equation (this is easily proved by differentiation 
of Eq. (\ref{cone}) with respect to coordinates $\{x_{A^\p A}\}$).  

	2. It gives rise to a point ``source'' $z^0_\mu(\sigma)$ 
in the complex space that ``influences'' at the point $X$. 

	3. If one knows $\sigma(X)$, twistor field of the 
congruence is immediately determined from the generating equations (\ref{rel}) 
and is constant in value along the elements of the complex null cone 
(\ref{cone}) connecting the ``source'' with the corresponding observation 
point $X$.

In the form (\ref{cone}), the caustics of the corresponding SFNC are 
evidently determined, instead of (\ref{sing}), by the requirement on 
the {\it retarded distance} to be null, 
\be{dist}
S^\p (\sigma): = \frac{dS}{d\sigma} = 0.
\ee
The {\it caustic locus equation} $S(X)=0$ follows then through elimination of 
$\sigma$ from (\ref{dist}) and substitution of the result into 
(\ref{cone}). It is easy to demonstrate 
that the obtained function $S(X)$ again satisfies the complex 
eikonal equation . 
 
To do that, let us differentiate, with respect to coordinates $X$, 
the complex null cone equation (\ref{cone}). Then one gets
\be{diff2}
\frac{1}{2} \nabl_{C^\p D}~S = Z_{C^\p D} - S^\p \nabl_{C^\p D}~\sigma.
\ee
By virtue of the caustic condition (\ref{dist}), the last term cancels, and 
one obtains
\be{eik}
\frac{1}{4} \nabl_{C^\p D} S \nabl^{C^\p D} S = Z_{C^\p D} Z^{C^\p D} = 0,
\ee
where one again makes use of the null cone equation (\ref{cone}).

Note that, remarkably, with any SFNC one can associate two functionally 
independent solutions of the complex eikonal equation. In the null cone 
representation  these are the ``retarded'' complex time $\sigma(X)$ 
and the caustic generating function $S(X)$. For the real case these 
properties are well known: both the ``field of retarded time'' of an 
arbitrary moving charge and the function of a ``field discontinuity'' 
surface do satisfy the (real) eikonal equation (see, e.g.,~\cite{Fock}). 
However, the principal spinor $g$ is a {\it complex} solution to the 
eikonal equation, even for the case of a charge moving in {\it real} space.

\section{The Kerr congruence and its generalization}

Recall that the well-known Kerr SFNC has been first obtained in the work
~\cite{Debney}. One may consider it as being generated by a virtual 
pointlike source {\it at rest} in complex extension ${\mathbb C}\bf M$, at 
a separation $a$ from the real slice ${\bf M}$. For this case, the 
null cone equation (\ref{cone}) takes the form 
\be{kerrcon}
S=(t-\sigma)^2-(z+\imath a)^2-\rho^2  = 0,~\rho:=\sqrt{x^2+y^2}, 
\ee
the caustics correspond to $\tau=\sigma$ (this is just the solution of the 
equation $S^\p = 0$) 
and, after substitution into (\ref{cone}) -- to the condition of {\it caustic 
locus}
\be{kerrsing}
-S = (z+\imath a)^2 + \rho^2 = 0,
\ee
which immediately results in the Kerr singular ring
\be{ring}
z=0,~~\rho = a .
\ee
The corresponding congruence is twofold, with rectilinear rays as  
elements of a family of hyperboloids, and the principal spinor $g$ is 
then determined  from Eqs. (\ref{rel}) as follows:
\be{spinor}
g= \frac{x+\imath y}{\wh z \pm \wh r}, ~~~\wh r = \sqrt{(z+\imath a)^2+\rho^2}. 
\ee
All this is well known and is presented here only for completeness of the 
exposition. For details related, in particular, to a possibility   
of generalizing the Kerr congruence in the way considered below we 
refer the reader to the old paper by Newman~\cite{Newman3} 

Let us consider now the case of a virtual charge ``moving'' uniformly  
in the complex space with ``imaginary'' velocity  $\imath v$, that is, let 
\be{imtraj}
z_0 = \sigma,~~z_3 = -\imath a+\imath v\sigma,~~z_1=z_2 = 0,
\ee
Then the generating null cone equation (\ref{cone}) aquires the form 
\be{imcone}
S=(t-\sigma)^2-(z+\imath a-\imath v\sigma)^2-\rho^2 = 0,
\ee
and the caustic condition reads 
\be{imcaust}
\frac{1}{2}S^\p=-(t-\sigma)+\imath v(z+\imath a-\imath v\sigma)=0
\ee
Resolving together Eqs.(\ref{imcone}) and (\ref{imcaust}), we find the shape 
and the dynamics of the caustic ring in the following form~\cite{gensol}: 
\be{imring}
z=0,~~~\rho= \left | -\frac{a}{\sqrt{1+v^2}}+\frac{v}{\sqrt{1+v^2}}t \right |. 
\ee
We thus see that the ring collapses into a point at the instant $t_0= a/v$ and 
then starts to expand, uniformly increasing its radius with the velocity 
\be{veloc}
V=\frac{v}{\sqrt{1+v^2}}, 
\ee
which is less that the fundamental one $V<c=1$, for {\it any} $v$, even 
for $v>1$.  This property 
is in correspondence with the old result of Bateman~\cite{Bateman} who 
had demonstrated that all of the (generally, stringlike) 
singularities of an arbitrary solution to the complex eikonal equation {\it lie on a 
surface propagating with a velocity which is always smaller than the 
fundamental one}. 

It is worth noting that the Kerr parameter $a$ loses its important role 
in the framework under consideration, marking only the instant at which the ring 
is contracted to a point. Remarkably, this  corresponds to the instant 
when the ``virtual'' charge pierces through the real physical slice $\bf M$ 
and continues afterwards its ``motion'' in the complex space extension 
${\mathbb C}\bf M$. 

The deformed SFNC itself is defined via its principal spinor $g$ from the  
twistor generating equations (\ref{rel}) and turns out to be of the form 
\be{defspinor}
g=(1+\imath v) \frac{x+\imath y}{(z-z_t) \pm \wh r},~~z_t:=-\imath a+\imath vt,
\ee
where the time-dependent complex ``distance'' from the virtual charge is 
\be{cdist}
\wh r = \sqrt{(z-z_t)^2+\rho^2(1+v^2)}.  
\ee
Notice that the branching points of the principal spinor $\wh r=0$ just 
correspond to the caustic ring (\ref{imring}). One can also make sure 
that the function (\ref{defspinor}) identically satisfies both the eikonal and 
the d'Alembert equations.  
	
Let us now calculate the strengths of electromagnetic field associated with 
the generalized Kerr congruence. To do that, one makes use of the representation 
(\ref{strength}) and of the particular form of the principal spinor 
(\ref{defspinor}). For the field strength components (in cylindrical 
coordinates) one the gets 
\be{cylindr}
{\cal E}_\rho = \pm \tilde e \frac{\rho}{\wh r^3},~~{\cal E}_z = \pm \tilde e
\frac{(z-z_t)}{\wh r^3},~~{\cal E}_\varphi = - v{\cal E}_\rho, 
\ee
where $\tilde e := e (1+v^2)$ while $\Re{\vec {\cal E}},\Im{\vec {\cal E}}$ 
correspond to the electric and magnetic field strengths, respectively.  

Formally, the obtained field exactly corresponds to that of a point charge 
moving uniformly in {\it real} space along $OZ$ with {\it imaginary} 
velocity. Since Maxwell's linear equations are insensitive to such a formal 
substitution of a complex entity, one is in fact able to obtain the above field 
from the very beginning, through the corresponding boost with an {\it imaginary} 
parameter, applied to the ordinary Coulomb field. In his old paper~\cite{Newman3},  
Newman has mentioned a possibility of obtaining such a solution by a complex 
boost. However, after separation of real (electric) and imaginary (magnetic) 
parts, the solution (\ref{cylindr}) looks indeed rather remarkable and 
possesses the following properties. 

1. The fields are two-valued, and the branches transform one into another 
under a continious path around the singular ring (just as it takes place 
in the case of the stationary Kerr field). Near the ring, they turn to infinity 
as $\sim 1/\delta^{3/2}$,~$\delta\to 0$ being the distance from the ring. 
	
2. The value of the electric charge $e$ (defined from (\ref{cylindr}) as the 
flow of he electric field through a surrounding 2-sphere) does not depend on 
time, on velocity; it is ``self-quantized'' (in correspondence with the general quantization 
theorem~\cite{sing04}) and equal to the ``elementary'' charge of the static 
(Coulomb) field.  The latter occures as the field of the fundamental static 
solution with $a=0,~v=0$,  for which the principal spinor $g$ represents the 
{\it stereographic projection} $S^2\mapsto \mathbb{C}$ (see, e.g.,
~\cite{AD,GR95,Asya}). (In the adopted dimensionless units, one has $e=\pm 1/4$). 

3. At the very instant when the singularity is pointlike, the field differs 
from the Coulomb one and possesses, in particular, {\it a vortex electric 
component} $E_\varphi$ which at large distances behaves as the radial 
Coulomb-like component, namely, as $\sim 1/r^2$; the magnetic field 
instantaneously vanishes. 

4. The most interesting case is, perhaps, that of the {\it ultrarelativistic} 
contraction/expansion of the singular ring, with the ``physical'' velocity 
$V\approx 1$ (this corresponds to the ``imaginary'' velocity of the virtual 
charge motion $v>>1$). Then at large distances $r>>r_0$,~~$r_0=V(t-t_0)$ 
being the current radius of the singular ring, both the vortex and radial 
components of the electric and magnetic fields are {\it compressed} along OZ 
and {\it concentrated in a sharp range of angles}, $\sin \theta\approx\sqrt{1-V^2}$. 
The same applies to the Pointing vector: it is non-null and may be great in 
absolute value in the above range of angles.

4. Generally, the magnetic moment, contrary to the electric charge, 
is variable; for a particular case of Kerr's stationary solution ($v=0$), 
it is known to be equal to $ea$, and in view of the value of mass and spin  
(defined by the metric of the Kerr-Newman solution), corresponds 
to the Dirac ``anomalous'' value. This fact has been first noticed by Carter
~\cite{Carter} and used further by Newman, Lopes, Burinskii et al. 
to ``model the electron'' in the framework of general relativity. At least for 
slow expansion/contraction, this remarkable property could be preserved 
(both the spin and the magnetic moment being proportional to the current 
radius of the singular ring); nonetheless, this conjecture should be 
verified by a corresponding exact solution to the Einstein-Maxwell set of 
equations, see below.   

\section{Discussion}

The existence of a time-dependent deformation of the Kerr congruence 
obtained in the paper seems to be a fact of fundamental importance. Indeed, 
an arbitrary small variation of the parameters of the initial static congruence 
qualitatively changes its character making it {\it unstable} with respect to 
its scale; as a result, the Kerr singular ring infinitely expands. In turn, 
the instability of the generating congruence lets one expect similar 
instability of the Kerr-Schild metric as the corresponding solution to the 
Einstein equations. 

In fact, there are obvious arguments in favour of the existence of such a 
solution: the general procedure of obtaining exact solutions of the (Maxwell-)Einstein 
(electro)vacuum equations starting from a shear-free congruence is well known 
(see, e.g.~\cite{Gursey}). 
Moreover, by analogy with the stationary Kerr case, for such a solution one can 
conjecture that, during contraction of the ``naked'' singular ring, a ``horizon'' 
would appear at some particular value of the ring current radius, so that one 
would actually have a continious {\it transition of the naked singularitiy into 
a rotating black hole}. On the other hand, during the ring's subsequent 
expansion, the ``black hole'' would throw off its horizon and become 
a ``naked'' singularity anew. We hope to prove and examine such solutions 
in a forthcoming publication. 

The observed instability also makes rather problematic the whole 
activity related to the possible ``particlelike'' 
interpretation of the Kerr singular ring (see, e.g.,~\cite{Burin,Newman2}). 
This becomes quite visual in  Newman's representation of a    
virtual charge ``moving'' in the ${\mathbb C}\bf M$ extension of   
Minkowski space-time. Actually, the singular (caustic) ring is formed together 
with the congruence of rays as the trace on $\bf M$ of the complex null cone of 
the charge. If we define the electromagnetic field via the congruence 
itself, as in the above presented scheme, the ring would automatically 
carry a unit electric charge and a magnetic moment with the right Dirac value 
of the gyromagnetic factor. However, the mass $M$ and the spin $J=Mca$ would be 
arbitrary and, to fix the ratio $J/M$, the virtual charge should ``oscillate'' 
in a  Compton-size ``complexified'' neighborhood of the real cut $\bf M$. 
However, neither in twistor theory nor in properties of the Einstein-Maxwell 
equations can one find any ground for such restrictions.

To overcome the instability related to the generically stringlike structure 
of singularities of any complex-valued field on $\bf M$, one can try to modify 
the very structure of the background physical space-time, say, by   
introduction of extra spacelike or timelike dimensions. However, in the 
framework of the above presented approach, it seems more natural to regard the 
complex space ${\mathbb C}\bf M$ itself in this capacity; then the transition 
to physical space-time could be realized via the {\it bilinear mapping} suggested 
in our work~\cite{Mink}.  The corresponding construction is motivated algebraically 
and, in phenomenological respect, naturally introduces an additional  
compact (phaselike) coordinate which allows for a geometrical 
explanation of the wave properties of particles~\cite{Pavlov07,AD09}. It is 
especially interesting that, in the paradigm of primordial complex 
space-time, the problem of quantization of characteristics of particlelike 
singularities is put off to a considerable extent. Indeed, on the 
$\mathbb{C}\bf M$ background, the fundamental equation of the complex light cone 
(\ref{cone}) has, generically, a great number of solutions $\{\sigma_N\}$ which  
fix the corresponding number of {\it observable} point singularities in 
distinct positions $\{z^0_N\}$. They all are {\it identical\/} in 
properties, being in fact one and the same particle at different positions on 
the {\it unique complexified world line}; in~\cite{Pavlov07} these have been 
called (an ensemble of) {\it duplicons\/}.

Finally, one can pass to studying, in the complex space-time background, 
the properties of a generic SFNC which, instead of a focal curve -- a world 
line of a virtual charge on the ${\mathbb C}\bf M$ space -- is generated by a 
{\it complex string}~\cite{gensol}. The arising structure of singularities 
seems to be rich enough and capable to describe {\it different types\/} 
of particlelike formations. We hope to classify them in forthcoming 
publications.  

\Acknow
It is a pleasure for me to thank Profs. G. Neugebauer and G. Sch\"{a}fer for  
interesting discussions and valuable remarks on the paper.

\end{document}